\documentclass[twocolumn]{emulateapj}

\usepackage{float}

\begin{document} 

\title{The Bright and Dark Sides of High-Redshift starburst galaxies from {\it Herschel} and {\it Subaru} observations}
\author{A. Puglisi\altaffilmark{1,2}} \author{E. Daddi\altaffilmark{3}} \author{A. Renzini\altaffilmark{4}} 
\author{G. Rodighiero\altaffilmark{1}} \author{J. D. Silverman\altaffilmark{5}} \author{D. Kashino\altaffilmark{6}} \author{L. Rodr{\'i}guez-Mu{\~n}oz\altaffilmark{1}} \author{C. Mancini\altaffilmark{1,4}} \author{V. Mainieri\altaffilmark{2}} \author{A. Man\altaffilmark{2}} \author{A. Franceschini\altaffilmark{1}} \author{F. Valentino\altaffilmark{3,11}} \author{A. Calabr{\`o} \altaffilmark{3}} \author{S. Jin\altaffilmark{3, 13}} \author{B. Darvish\altaffilmark{12}} \author{C. Maier\altaffilmark{7}} \author{J. S. Kartaltepe\altaffilmark{8,9}} \author{D. B. Sanders\altaffilmark{10}}
\affiliation{1, Dipartimento di Fisica e Astronomia, Universit{\`a} di Padova, vicolo dell’Osservatorio 2, 35122 Padova, Italy
\\2, ESO, Karl-Schwarschild-Stra{\ss}e 2, 85748 Garching bei M{\"u}nchen, Germany
\\3, Laboratoire AIM-Paris-Saclay, CEA/DSM-CNRS-Université Paris Diderot, Irfu/Service d’Astrophysique, CEA Saclay, Orme des Merisiers, 91191 Gif sur Yvette, France
\\4,INAF-Osservatorio Astronomico di Padova, Vicolo dell’Osservatorio, 5, 35122 Padova, Italy
\\5, Kavli Institute for the Physics and Mathematics of the Universe (WPI), Todai Institutes for 
for Advanced Study, the University of Tokyo, Kashiwanoha, Kashiwa, 277-8583, Japan
\\6, Institute for Astronomy, Department of Physics, ETH Z{\"u}rich, Wolfgang-Pauli-strasse 27, CH-8093 Z{\"u}rich, Switzerland
\\7, University of Vienna, Department of Astrophysics, Tuerkenschanzstrasse 17, 1180 Vienna, Austria
\\8, National Optical Astronomy Observatory, 950N. Cherry Ave, Tucson AZ 85719, USA
\\9, School of Physics and Astronomy, Rochester Institute of Technology, 84 Lomb Memorial Dr., Rochester, NY 14623, USA
\\10, Institute for Astronomy, University of Hawaii, 2680 Woodlawn Drive, Honolulu, HI 96822, USA
\\11, Dark Cosmology Centre, Niels Bohr Institute, University of Copenhagen, Juliane Mariesvej 30, DK-2100 Copenhagen, Denmark
\\12, Cahill Center for Astrophysics, California Institute of Technology, 1216 East California Boulevard
Pasadena, CA 91125
\\13, Key Laboratory of Modern Astronomy and Astrophysics in Ministry of Education, School of Astronomy and Space Science, Nanjing University, Nanjing 210093, China 
}

\begin{abstract}
We present rest-frame optical spectra from the FMOS-COSMOS survey of twelve $z \sim 1.6$ \textit{Herschel} starburst galaxies, with Star Formation Rate (SFR) elevated by $\times$8, on average, above the star-forming Main Sequence (MS). Comparing the H$\alpha$ to IR luminosity ratio and the Balmer Decrement we find that the optically-thin regions of the sources contain on average only $\sim 10$ percent of the total SFR whereas $\sim90$ percent comes from an extremely obscured component which is revealed only by far-IR observations and is optically-thick even in H$\alpha$.
We measure the [NII]$_{6583}$/H$\alpha$ ratio, suggesting that the less obscured regions have a metal content similar to that of the MS population at the same stellar masses and redshifts. However, our objects appear to be metal-rich outliers from the metallicity-SFR anticorrelation observed at fixed stellar mass for the MS population.
The [SII]$_{6732}$/[SII]$_{6717}$ ratio from the average spectrum indicates an electron density $n_{\rm e} \sim 1,100\ \mathrm{cm}^{-3}$, larger than what estimated for MS galaxies but only at the 1.5$\sigma$ level.
Our results provide supporting evidence that high-$z$ MS outliers are the analogous of local ULIRGs, and are consistent with a major merger origin for the starburst event. 
\end{abstract}

\section{Introduction} 
\label{intro}
The majority of Star Forming (SF) galaxies at all redshifts form stars in a quasi-steady state along the Main Sequence (MS), a correlation between their stellar mass (M$_{\star}$) and the Star Formation Rate \citep[SFR,][]{Noeske07,Elbaz07, RenziniPeng}. 
MS galaxies also form a tight sequence in the M$_{\star}$-metallicity (Z) plane \citep[the Mass-Metallicity Relation MZR, see e.g.,][]{Tremonti04, Maiolino08}, with metallicity increasing with M$_{\star}$. 
The scatter in the MZR is reduced when considering the anti-correlation of Z with the SFR at fixed M$_{\star}$ \citep[e.g.][]{Ellison08}. \cite{Mannucci10} found such relation to be invariant up to $z \sim 2.5$ and called it the Fundamental Metallicity Relation (FMR). The interpretation of this relation is that the SFR is enhanced by upward fluctuations in the gas inflow rate from the cosmic web, while such inflow dilutes the metal content of the system \citep[e.g.,][]{Lilly13}. 
A fourth key parameter of SF galaxies is their dust content, which correlates with their M$_{\star}$, SFR and Z so that more massive objects, as well as most SF or metal-rich galaxies tend to host larger dust reservoirs, thus suffering higher extinction \citep{GarnBest, Pannella14, Tan14}.

Besides the MS galaxy population, a population of outliers has been observed at all redshifts, supporting extreme SFRs for their M$_{\star}$. While such MS-outliers contribute only $\sim 10 \%$ to the cosmic Star Formation History (SFH) \citep{Rodighiero11},  they may represent a key phase in galaxy evolution, having been considered among the progenitors of passively evolving ellipticals \citep[e.g.][]{Cimatti08}.
Local outliers appear to be mainly Ultra-Luminous InfraRed Galaxies (ULIRGs) undergoing a major merger that funnels gas into the nucleus, enhances the Star Formation Efficiency (SFE) and triggers a StarBurst \citep[SB, ][]{SandersMirabel}.
These systems have strong dust extinction \citep{MonrealIbero10} and complex kinematic and gas properties \citep{Rich15Goals}, with strong nuclear outflows and shock dominated regions \citep{Westmoquette09}.
The nature of the high-$z$ counterparts to these SB galaxies is however still debated, as it is not clear yet to which extent their high SFR is uniquely due to a higher SFE, as if experiencing a different mode of SF \citep[e.g.][]{daddi10,tacconi13,tacconi17, sargent14, Silverman15}, or to a higher gas content, or a combination thereof \citep{Genzel15,Scoville16}. 
Studies of the metal content of high-$z$ SBs may shed light on their nature as the gas-phase metallicity reflects their recent SF activity and is a crucial input when estimating their gas content via either the CO luminosity or the dust continuum emission \citep[e.g.][]{Genzel15, tacconi17}.

In this work we present an analysis of the Inter Stellar Medium (ISM) properties of 12 \textit{Herschel}-selected SB galaxies at $1.4 \leq z_{\rm spec} \leq 1.7$ with H$\alpha$ detection via near-IR spectroscopy from the FMOS-COSMOS survey \citep{Silverman14}. These galaxies have SFRs from $4 \ \rm{ to \ over} \ 10 \times {\rm SFR}_{\rm MS}$ and are analyzed in comparison to the MS population, taking advantages of our complementary studies on MS galaxies at the same redshift \citep{Zahid14, Kashino13, Kashino16}.
Throughout this Letter we adopt a \citet{Chabrier} IMF, standard cosmology ($H_{0}=70 \rm km s^{-1}Mpc^{-1}, \Omega_{\rm m} = 0.3, \Omega_\Lambda = 0.7$), AB magnitudes and a \cite{Calzetti00} extinction law.

\section{Data set and sample selection}
\label{sec:sample}
\begin{figure}
\begin{center}

\centering
\epsscale{1.2}
\figurenum{1}
\plotone{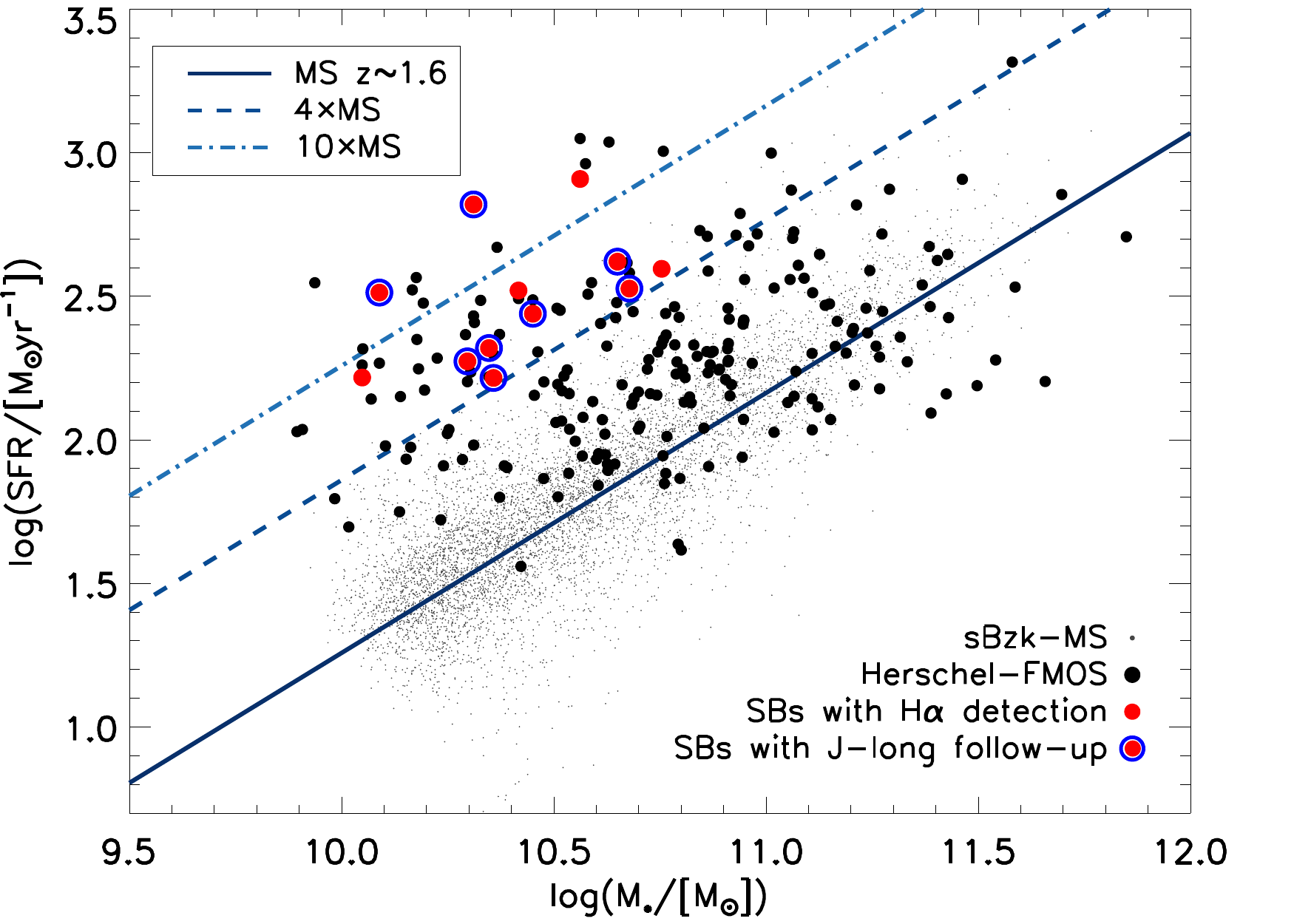}
\caption{SFR as a function of M$_{\star}$ for the \textit{Herschel}-FMOS sources. Small grey dots are the sBzk sources used to define the MS.
Red filled circles indicate the \textit{Herschel} SBs with H$\alpha$ detection. Those with a \textit{J}-long follow-up are highlighted with blue circles. 
\label{fig:MS}}
\end{center}
\end{figure}

SB galaxies are identified for lying above the MS, hence a careful definition of the MS at the redshift of interest is required.
We defined the MS equation (${{\rm log(SFR} / {[M}_{\odot}\mathrm{/yr])}} = 0.906 \times \mathrm{log}(M_{\star}{/[M}_{\odot}\mathrm{])} - 7.798 $, blue solid line in Fig. \ref{fig:MS}) from a sample of $\sim 3000$ star-forming Bzk galaxies \citep{Daddi04} at $1.4 \leq z_{\rm phot} \leq 1.7$ in the COSMOS field, for which we computed M$_{\star}$ and SFR by fitting their Spectral Energy Distribution (SED), using broad-band photometry from the \cite{Laigle16} catalog and the \textit{hyperzmass} code \citep{Bolzonella} with \cite{BC03} stellar populations synthesis models and constant SFH. 

\begin{figure}
\epsscale{1.2}
\figurenum{2}
\plotone{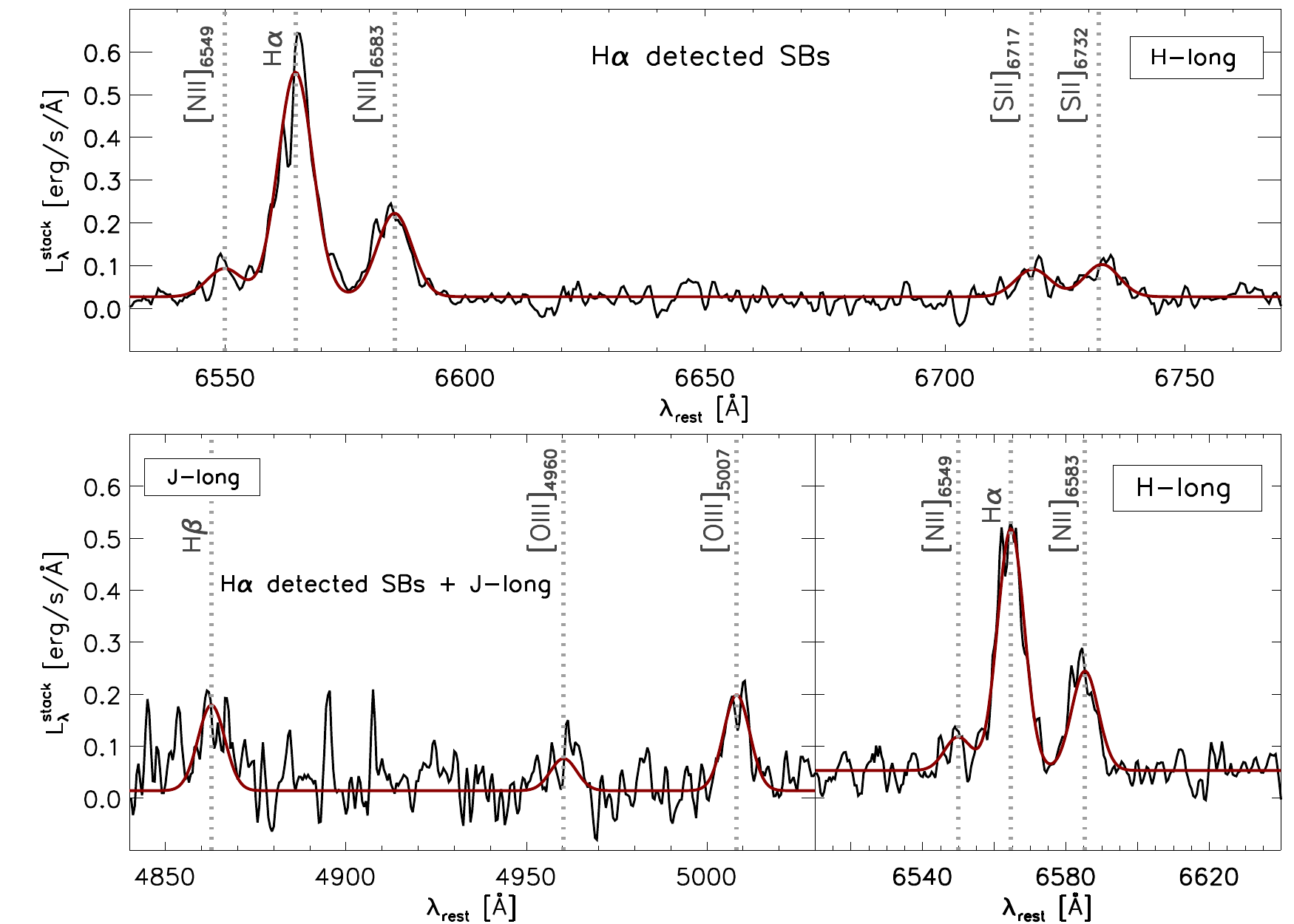}
\caption{\textit{Upper panel:} average \textit{H}-long spectrum of the 12 H$\alpha$-detected SB sources.
\textit{Lower panel:} average \textit{J}-long (\textit{left}) and \textit{H}-long (\textit{right}) spectra for the 8 H$\alpha$-detected SBs with \textit{J-long} follow-up. 
The red curve is the emission lines + continuum fit.
\label{fig:Average Spectrum}}
\end{figure}

Our SBs are drawn from a \textit{Herschel} sample (see \citealt{Rodighiero11} for details about the PACS photometry) at $1.4 \leqslant z_{\rm phot} \leqslant 1.7$ having near-IR spectroscopy from the FMOS-COSMOS survey. FMOS observations with the \textit{H}-long and \textit{J}-long gratings allows us to detect H$\alpha$, [NII]$_{6549, 6583}$ and H$\beta$,[OIII]$_{4959, 5007}$ emission lines, respectively. \cite{Kashino13, Kashino16} and \cite{Silverman14} contain further details about the spectroscopic observations and the data analysis. We also refer the reader to our companion papers for the description of emission line fitting and flux measurements, as well as the stacking technique adopted to construct average spectra as shown in Fig. \ref{fig:Average Spectrum}.

SB sources analyzed in this work are selected as outliers from the MS defined above, with $\rm SFR \geq 4 \times \rm SFR_{MS}$ (see Fig. \ref{fig:MS}), following \cite{Rodighiero11}. Our selection may overlap with samples of Sub-mm selected Galaxies (SMGs). However, SMGs galaxies represent a mixed population \citep{Rodighiero11, Roseboom13} that include both massive MS galaxies as well as objects qualified as of SBs according to our criterion. Instead, our criterion selects a pure sample of SB galaxies by construction.

For each \textit{Herschel}-FMOS source, we compute $M_{\star}$ using homogeneous photometry and methodology as applied for MS galaxies, while fixing the redshift to the FMOS $z_{\rm spec}$.
SFRs are measured from the bolometric IR luminosity ($L_{\rm FIR}$), using the calibration of \cite{KSFR98}, rescaled to a Chabrier IMF by dividing by a factor of 1.7.  The $L_{\rm FIR}$ are computed by integrating the far-IR SED over the range $\lambda \in [8-1000] \mu$m, with the SEDs being derived  by fitting the \cite{Magdis12} templates to the \textit{Herschel} photometry (PACS 100 and 160 $\mu$m, SPIRE 250, 350 and 500 $\mu$m).

AGN candidates among the SBs are identified and discarded using the BPT diagram and the \cite{Kewley13} dividing line at $z \sim 1.6$ (when H$\alpha$, H$\beta$, [OIII]$_{5007}$ and [NII]$_{6583}$ are detected), the [NII]$_{6583}/\rm H\alpha$ ratio (if only [NII]$_{6583}$ and H$\alpha$ are available), X-ray detections and inspection of the mid-IR SED (Marcella Brusa, private communication).
With these criteria we identify 8 AGN, resulting in an AGN fraction among the off-MS sample of $\sim \ 40$\%, substantially higher than the $\sim \ 8$\% identified on the MS by \cite{Kashino16}.
The average spectrum of these AGN candidates differs from the one of the ``purely SF" population, showing broader H$\alpha$ and [NII]$_{6549, 6583}$ emissions and an additional broad component on the H$\alpha$ emission.

The sample analyzed here includes 12 H$\alpha$-detected SBs (red filled circles in Figure \ref{fig:MS}, upper panel of Fig. \ref{fig:Average Spectrum}).
Table \ref{tab:bins_properties} summarizes the main physical properties of these sources.  Among them, 8 objects have a \textit{J}-long follow-up (blue circles in Fig. \ref{fig:MS}, lower panel of Fig. \ref{fig:Average Spectrum}). 
Seven of these objects have molecular gas measurements through ALMA CO 2-1 observations, indicating an average gas fraction $f_{\rm gas} \sim 50 \%$, which implies very high SFE and short depletion timescale \citep[$\sim 40$ Myrs]{Silverman15}.

\begin{deluxetable*}{cccccccc}
\tablecolumns{8} 
\tablewidth{0pc} 
\tablecaption{Physical properties of the $z \sim 1.6$ SB sample. } 
\tablehead{ 
\colhead{PACS-ID} & \colhead{RA}   & \colhead{Dec}    & \colhead{z$_{spec}$} & 
\colhead{log(M$_{\star})$}    & \colhead{SFR$_{\mathrm{\rm FIR}}$} $\pm 1\sigma$  & \colhead{F(H$\alpha$)  $\pm 1\sigma$}    & \colhead{F([NII]$_{6583}$) $\pm 1\sigma$} \\
\phn & \colhead{hours}   & \colhead{deg}    & \colhead{} & 
\colhead{M$_{\odot}$}    & \colhead{M$_{\odot}yr^{-1}$} & \colhead{$10^{-16} erg s^{-1} cm^{-2}$} &\colhead{$10^{-16} erg s^{-1} cm^{-2}$}} \\
\cline{1-8}
\startdata 
  300 & 09:58:24.32 & 2:15:15.10 & 1.6706 & 10.4  & 332 $\pm$ 8 & 1.88 $\pm$ 0.04 & 0.38 $\pm$  0.03\\
  299 & 09:59:41.31 & 2:14:42.80 & 1.6467 & 10.1  & 326 $\pm$ 14 & 1.24 $\pm$ 0.32 & 0.43 $\pm$  0.26 \\
  455 & 09:59:43.88 & 2:38:08.20  & 1.6696 & 10.4  & 275 $\pm$ 11 & 1.45 $\pm$ 0.12 & 0.39 $\pm$  0.06\\
  491 & 10:00:05.19 & 2:42:04.70  & 1.6366 & 10.4   & 209 $\pm$ 21 & 1.15 $\pm$ 0.08 & 0.35 $\pm$  0.06\\
  830 & 10:00:08.73 & 2:19:02.50  & 1.4610 & 10.7   & 336 $\pm$ 34 & 1.53 $\pm$ 0.09 & 0.41 $\pm$  0.05\\
  472 & 10:00:08.95 & 2:40:10.50 & 1.5988 & 10.3  & 661 $\pm$  66 & 0.78 $\pm$ 0.04 & 0.24 $\pm$  0.04\\
  135 & 10:00:15.72 & 1:49:48.00 & 1.6508 & 10.3   & 188 $\pm$  6 & 0.46 $\pm$ 0.19 & 0.15 $\pm$  0.16\\
  175 & 10:00:34.62 & 1:55:25.40 & 1.6664 & 10.4   & 164 $\pm$  9 & 1.32 $\pm$ 0.07 & 0.30 $\pm$  0.03\\
  682 & 10:01:23.96 & 1:52:28.60  & 1.4681 & 10.6   & 418 $\pm$ 8  & 1.21 $\pm$ 0.19 & 0.58 $\pm$  0.15\\
  197 & 10:01:34.46 & 1:58:47.70 & 1.6005 & 10.7   & 394 $\pm$ 11 & 0.82 $\pm$ 0.07 & 0.32 $\pm$  0.08\\
  787 & 10:02:27.95 & 2:10:04.70  & 1.5234 & 10.6  & 811 $\pm$ 81 & 1.42 $\pm$ 0.05 & 0.75 $\pm$  0.03\\
  251 & 10:02:39.64 & 2:08:47.10 & 1.5847 & 10.0   & 165 $\pm$ 16 & 2.71 $\pm$ 0.07 & 0.59 $\pm$  0.04\\
\enddata 
 \tablecomments{The $1 \sigma$ error on M$_{*}$ is 0.1 $dex$. Emission line fluxes are aperture corrected, as described in \cite{Silverman14}. Absolute line flux values are also affected by the error on the aperture correction \citep[0.17 $dex$, ][]{Silverman14}}
\label{tab:bins_properties}
\end{deluxetable*}

\section{Results}
\label{results}

The key physical quantities examined in the following are measured from different portions of the rest-frame spectrum: M$_{\star}$ is primary sampled by the near-IR photometry, the SFR by the far-IR luminosity, while the gas-phase metallicity and the electron density from the optical spectrum.
Given the complex morphology and the high extinction of SB galaxies, each indicator may not trace the same component in each system and we carefully assess which possible galaxy component the various measurements likely refer to.

\begin{figure}
\figurenum{3}
\centering
\epsscale{1.2}
\plotone{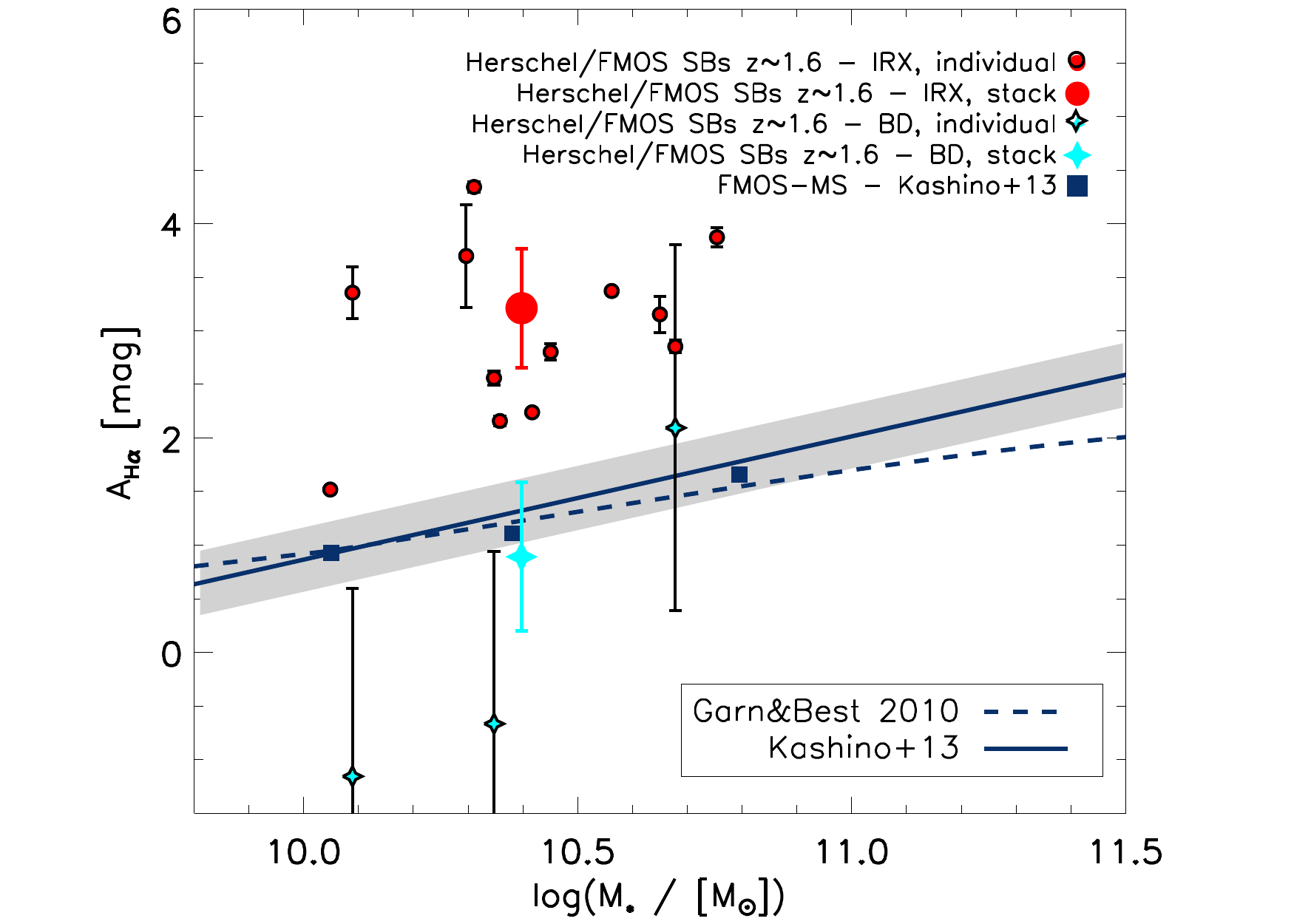}
\caption{H$\alpha$ attenuation as a function of M$_{\star}$ for $z \sim 1.6$ SBs, derived from the BD and the IRX. Our measurements are compared with the MS-trend at $z \sim 1.6$ and in the local Universe (solid and dashed lines respectively). 
\label{fig:AHA_MSTAR}}
\end{figure}

We use two methods for estimating the nebular attenuation A$_{H\alpha}$ in our sample: the H$\alpha$/H$\beta$ emission line ratio (the Balmer Decrement, BD), assuming Case B recombination and a gas temperature $T = 10^4 \rm K$:
\begin{equation}
A_{\rm H\alpha, BD} = 	\frac{2.5}{k_{\rm H\beta} - k_{\rm H\alpha}}\mathrm{log}[\frac{H\alpha/H\beta}{2.86}] \times k_{\rm H\alpha}
\end{equation} 
and the SFR$_{\rm FIR}$/SFR$_{\rm H\alpha,\rm \ obs}$ ratio (the {\it IRX} ratio):
\begin{equation}
A_{H\alpha, IRX} = 2.5\mathrm{log} \times (1 + \frac{\mathrm{SFR}_{\rm FIR}}{\mathrm{SFR}_{\rm H\alpha, obs}}). 
\end{equation} 
From our average spectrum we obtain $A_{\rm H\alpha, BD} \simeq 0.89 \pm 0.69 \ \mathrm{mag}$, while the IRX method gives $A_{\rm H\alpha, IRX} \sim 3.3 \  \pm 0.51 \ \mathrm{mag}$, much larger than the former and than the typical MS value, as seen in Fig. \ref{fig:AHA_MSTAR}.

This means that the optical light (from which A$_{\rm H\alpha, BD}$ is measured) comes from relatively unobscured lines of sight, whereas the bulk of the SF is almost completely obscured at optical wavelengths. 
This is commonly observed in local ULIRGs where the starburst usually takes place in a region not coincident with optical bright regions or dark dust lanes observed in the optical \citep[e.g., ][]{Poggianti01} and does not contribute to the optical emission, while it dominates the bolometric energy output of the system with the FIR emission.

Quantifying how much the heavily obscured and optically bright components contribute to our spatially unresolved measurements (both the optical spectrum and the FIR emission) is key for interpreting our results. 
To asses the amount of nebular emission arising from the obscured core, we model our average galaxy with two components: an optically thin part (with intrinsic nebular emission L$_{\rm H\alpha, thin}$ and attenuation A$_{\rm H\alpha, thin}$) and an optically thick region (characterized by L$_{\rm H\alpha, thick}$ and A$_{\rm H\alpha, thick}$) corresponding to the starburst core. 
The total SFR of the system is given by the sum of SFR$_{\rm FIR}$, obtained from the average L$_{\rm FIR}$ of our sample, and SFR$_{\rm H\alpha, stack}$, measured from the stacked spectrum and uncorrected for dust attenuation, thus SFR$_{\rm tot} = (322 + 16.2) = 338.2 \ M_{\odot}yr^{-1}$. 
Both the thick and thin component will contribute to SFR$_{\rm tot}$ and to our integrated measurements, with their relative contribution depending on their attenuation. We cannot quantify the value of A$_{\rm H\alpha, thin}$ and A$_{\rm H\alpha, thick}$ with our data, but we can solve the problem in two extreme cases, bracketing a range of possible configurations. 
For the first limiting case, we assume a fully obscured thick core (A$_{\rm H\alpha, thick} = \infty$) which will not contribute to the optical emission. Therefore, the BD gives the extinction of the thin component only, that has a dust-corrected SFR$_{\rm H\alpha, thin} = 36.8 M_{\odot}yr^{-1}$ and will partially contribute to the FIR emission. As such, the contribution of the thick component to SFR$_{\rm FIR}$ will be SFR$_{\rm FIR, thick}$ = SFR$_{\rm FIR}$ - SFR$_{\rm FIR, thin}$ = 301.4 $M_{\odot}yr^{-1}$ or $\sim 89$ \% of SFR$_{\rm tot}$. 
The other extreme situation is the one in which the thin component is dust free (A$_{\rm H\alpha, thin} = 0$) thus it will not contribute to the FIR emission, which samples the thick core only. In this case the intrinsic H$\alpha$ emission from the thick core can be inferred from SFR$_{\rm FIR}$ while its attenuation A$_{\rm H\alpha, thick}$ can be derived as the minimum attenuation required on L$_{H\alpha, thick}$ to obtain an attenuation-free thin component. With this computation, we estimate a lower limit for the attenuation of the thick core A$_{\rm H\alpha, thick} = 4.5$ mag, which would imply that the H$\alpha$ emission from the core is attenuated by a factor of at least $\sim$ 70, being still optically thick. 
In this limiting situation the thick component has its maximum possible contribution to the optical emission lines of $\sim \ 33$\%,  while contributing $\sim$ 94\% of the SFR$_{\rm tot}$.

\begin{figure*}
\figurenum{4}
\centering
    \epsscale{1}
    \plotone{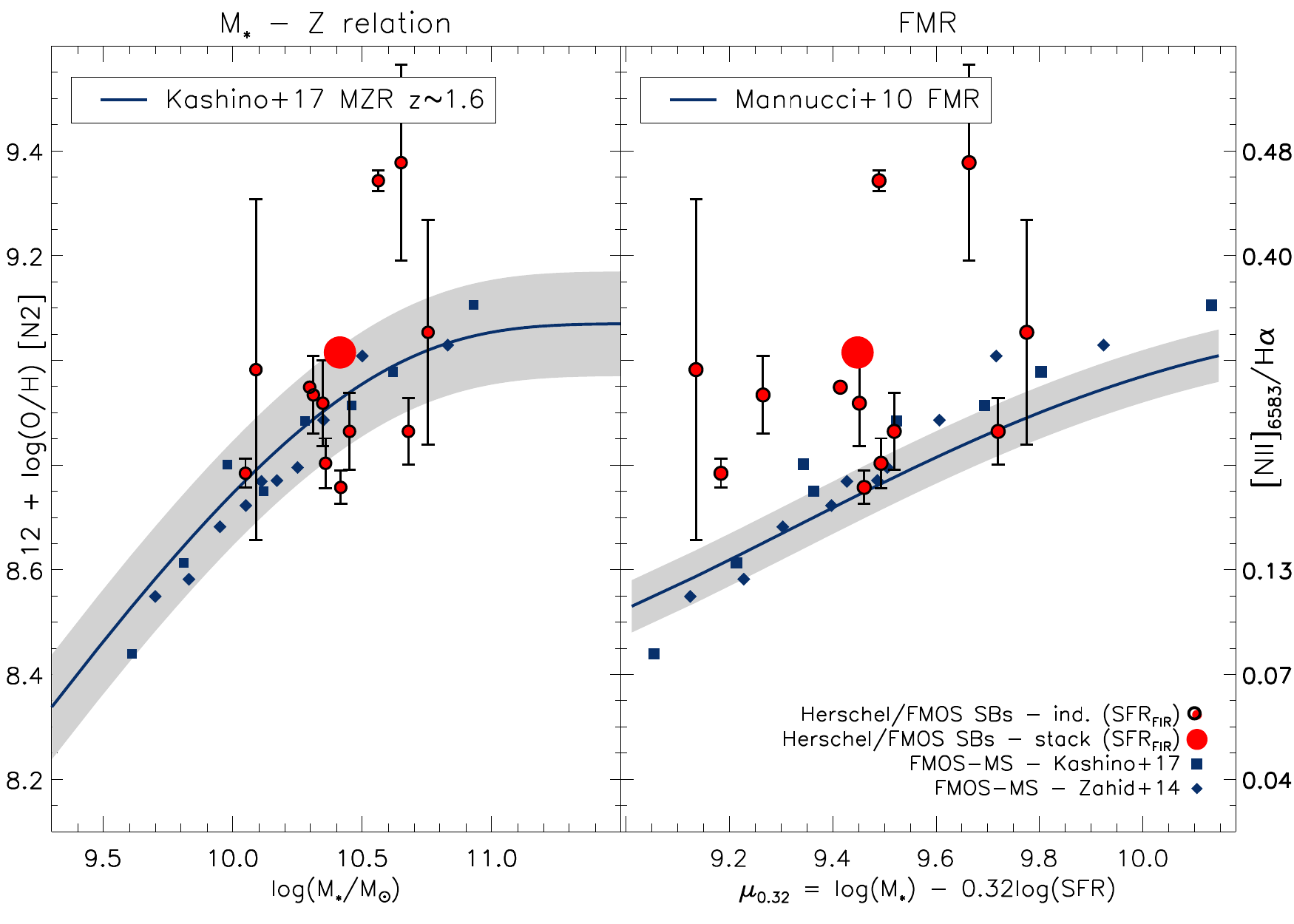}
\caption{
\textit{Left:} Metallicity as a function of M$_{\star}$ for the SB sample based on individual and stacked spectra, compared with the MZR and data points at $z \sim 1.6$ from the FMOS-MS sample of \cite{Kashino16} and \cite{Zahid14}. The shaded area marks a range of $\sim 0.1 \ dex$, roughly the scatter of this relation. 
\textit{Right:} FMR for SB galaxies compared with the \cite{Mannucci10} equation and its scatter of $\sim 0.05 \ dex$. The color code is as in the left panel. The right axis indicates the [NII]$_{6583} / \rm H \alpha$ ratio.}
\label{fig:Metallicity}
\end{figure*}

Having in mind the caveat that the emission-line ratios may not be representative of the optically thick core, we measure the metallicity (O/H) from  the [NII]$_{6583}$/H$\alpha$ ratio, adopting the \cite{Maiolino08} calibration for both the SB and MS samples.
In the left panel of Figure \ref{fig:Metallicity} we show the MZR for our SBs, compared with the trend from \cite{Kashino16} and \cite{Zahid14}.
Most of our individual estimates are consistent with the MS trend, although with some scatter. We note in particular two high-metallicity outliers, possibly late-stage mergers enriched by the SB itself, or sources with a low-level AGN emission, as their position in the BPT diagram is relatively close to the dividing line.
However, the estimate from the stacked spectrum indicates that (on average) the metallicity of SBs is consistent with that of MS galaxies.

We explore the correlation between O/H, $M_{\star}$ and SFR in the right panel of Fig. \ref{fig:Metallicity}: SBs appear as outliers in this plot, showing a metal content significantly higher than expected from the FMR, which comes directly from them having near-MS metallicities but much higher SFR.

We note that the [NII]$_{6583}$/H$\alpha$ ratio can be contaminated by shocks and this contribution may be important in SB galaxies \citep[see, e.g.,][]{Westmoquette09}. 
Still, the location of our sample in the [SII]-BPT diagram (constructed from their average spectrum, shown in Fig. \ref{fig:Average Spectrum}) confirms that their line ratios are dominated by SF.
Moreover, our conclusion regarding the average metallicity based on the calibration of \cite{Dopita16} (independent of the ISM pressure and ionization parameter) remain unaltered. 

Lastly, we use the [SII] lines to measure the electron density $n_{\rm e}$ for our SBs from their average spectrum. We convert the ratio [SII]$_{6717}$/[SII]$_{6732}$ to $n_{\rm e}$ using the \textit{temden} package in IRAF, assuming an electron temperature $T_{\rm e} = 10,000$ K, which is typical for HII regions. The average electron density derived for our 12 SBs is $n_{\rm e} = 1,111^{+1143}_{-587} \mathrm{cm}^{-3}$ ([SII]$_{6717}$/[SII]$_{6732}$ = 0.851 $\pm$ 0.167), higher than the average $n_{\rm e} \sim 220 ^{+172}_{-128}\mathrm{cm}^{-3}$ ([SII]$_{6717}$/[SII]$_{6732}$ = 1.213 $\pm$ 0.11) measured in the same way on the MS by \cite{Kashino16}, but only by 1.5$\sigma$.

\section{Discussion and conclusions}
 
\label{Conclusions}
Different recipes fail to recover the nebular attenuation in our SBs, as the BD gives an attenuation almost 4$\times$ smaller than what measured from the far-IR, resulting in an attenuation-corrected SFR$_{\rm H\alpha}$ that is $\sim 10\%$ of SFR$_{\rm tot}$. This is at odds with observations on the MS, where different attenuation diagnostics fairly agree \citep[][]{Kashino13, Rodighiero14, Puglisi16, Shivaei16} and leads to describe SBs as composed of two components: a compact, optically thick core, and an optically thin part, that primarily contributes in the optical continuum and line emissions.
Measurements of the electron density through the [SII] doublet is fairly high, indicating that even the optically thin part is denser than in normal MS galaxies and perhaps suggests that we might be starting to probe the denser starburst core, similar to local SBs \citep{Juneau09}. This result will require further confirmation at higher S/N.
A visual inspection of the ALMA CO vs optical morphology independently supports our interpretation as, at least in few cases, the peak of the CO emission tracing the molecular gas corresponds to a region undetected at optical wavelengths while IRAC re-aligns with the ALMA emission \citep[see Fig. 2 in ][]{Silverman15}, in a similar way to what is observed for some high-$z$ SMGs \citep{Simpson16}.
However, comparisons between the BD and IRX in high-z SMGs show that these two diagnostics are in agreement \citep{Takata06}, at odds with our results. This is expected, as the SMGs population includes both massive MS galaxies and off-MS sources.

The complex structure of our sources is revealed also from their average FWHM(H$\alpha) \sim 390 \ km/s$, higher than the MS estimate at similar M$_{\star}$ (313 $km/s$ at log(M$_{\star}) \sim 10.48$, see Fig. 2 in \citealt{Kashino16}): this enhancement might indicate the presence of tidal fields, merger signatures and/or outflows in the optically-thin regions of our SBs.

Our results disfavor the possibility that SBs are simply isolated extremely gas-rich galaxies, while supporting a merger-driven scenario for their origin, as the latter scenario can lead to the formation of a heavily obscured compact core in the center. 
A caveat here is that the estimate of A$_{H\alpha}$ involves using aperture corrections computed from I$_{\rm F814W}$-band photometry to account for the flux falling outside the fiber \citep[see Sect. 7.2 in][]{Silverman14}. One might wonder that for such peculiar objects the continuum and emission line morphologies may strongly differ, unlike for the MS. 
Nevertheless, the attenuation at H$\alpha$ and observed I-band continuum (hence about 2800 {\AA} at $z\sim1.6$) are expected to be nearly identical \citep[when accounting for the extra attenuation of the line, e.g.][]{Kashino13}.
Also, the tension between different attenuation diagnostics persists even when aperture corrections are not applied, thus not altering our conclusions. 
Future integral field observations can help to overcome these uncertainties and also to recover the effective nebular extinction pattern, that may be patchy. 

Finally, we note that the BD-IRX mismatch could also arise by the fact that FIR samples the SFR on 
a longer timescale than H$\alpha$. Such an effect would be detectable only on short (several tens of Myr) timescales and only in case of strong SFR variations. Any such trend would be averaged out in our average measurements.

We also add another important piece to our understanding of the high-$z$ SB physics. While ``normal'' galaxies follow a correlation in the M$_{\star}$-Z-SFR plane, our SBs are outliers from the FMR, due to a metal-content substantially higher than one would predict from their M$_{\star}$ and SFR. 
This is somewhat expected if the core is thick at $\lambda_{\rm H\alpha}$,  but it further supports the idea that the starburst would not be triggered by a \textit{major} accretion event of metal-poor gas, because such event would have caused a ISM dilution, reducing its metal content also presumably in its surroundings.
However, we must note that the contribution of the core to the optical spectrum might reach at most 30\% (likely, it is much lower) and our measurement mainly refers to the thin component. Still, this metallicity may be a lower limit to the metal-content of the heavily obscured core, if the ongoing SF has already significantly enriched the bulk of the ISM even if there are evidences in local ULIRG that the metallicity is uniform because of rapid mixing \citep{Vais17}.
These local studies, however, make use of optical emission lines indicators thus not probing still the metal content of the heavily obscured core. Given that the starburst region is optically thick, near-IR rest-frame or sub-mm observations would be needed to measure its metal content. 

In conclusion, our work demonstrates for the first time that high-$z$ SB galaxies behave like local ULIRGs, with a heavily obscured, probably metal-rich core that hosts $\gtrsim$90 \% of their vigorous SF, and supports the scenario in which high-z SBs are experiencing a violent and rapid SF episode because of their high density gas concentrations, probably driven by a major merger as in the local Universe. 

\section{Acknowledgements}
A.R. acknowledges partial support by the DFG Cluster of Excellence ʻOrigin and Structure of the Universeʼ (www.universe-cluster.de). B.D. acknowledges financial support from NASA through the Astrophysics Data Analysis Program (ADAP), grant number NNX12AE20G.
We thank the referee for the careful reading of the manuscript and for giving valuable comments and insights.
The author thanks Rob Ivison for useful discussions and Margherita Talia for reading the manuscript and providing useful comments.
The author thanks also Suhail Dhawan and Miguel Querejeta for helpful discussions and support while this paper was completed. 
\bibliographystyle{aasjournal}

\end{document}